# A Self-assessment Instrument for Assessing Test Automation Maturity

Yuqing Wang
M3S, ITEE, University of Oulu
Oulu, Finland
yuqing.wang@oulu.fi

Mika Mäntylä
M3S, ITEE, University of Oulu
Oulu, Finland
mika.mantyla@oulu.fi

Sigrid Eldh
Ericsson AB
Stockholm, Sweden
sigrid.eldh@ericsson.com

Jouni Markkula
M3S, ITEE, University of Oulu
Oulu, Finland
jouni.markkula@oulu.fi

Kristian Wiklund
Ericsson AB
Stockholm, Sweden
kristian.wiklund@ericsson.com

Tatu Kairi
Eficode
Helsinki, Finland
tatu.kairi@eficode.com

Päivi Raulamo-Jurvanen
M3S, ITEE, University of Oulu
Oulu, Finland
paivi.raulamo-jurvanen@oulu.fi

Antti Haukinen
Comiq
Helsinki, Finland
antti.haukinen@comiq.fi

## ABSTRACT

Test automation is important in the software industry but self-assessment instruments for assessing its maturity are not sufficient. The two objectives of this study are to synthesize what an organization should focus to assess its test automation; develop a self-assessment instrument (a survey) for assessing test automation maturity and scientifically evaluate it. We carried out the study in four stages. First, a literature review of 25 sources was conducted. Second, the initial instrument was developed. Third, seven experts from five companies evaluated the initial instrument. Content Validity Index and Cognitive Interview methods were used. Fourth, we revised the developed instrument. Our contributions are as follows: (a) we collected practices mapped into 15 key areas that indicate where an organization should focus to assess its test automation; (b) we developed and evaluated a self-assessment instrument for assessing test automation maturity; (c) we discuss important topics such as response bias that threatens self-assessment instruments. Our results help companies and researchers to understand and improve test automation practices and processes.

## 1 INTRODUCTION

The software industry has been focusing on the adaption of Agile development and more recently on Continuous Delivery capabilities. It is widely seen that these lead to faster time-to-market, better satisfaction of clients, and higher product quality [9]. However, in Agile software development manual testing is a bottleneck preventing efficient, rapid, reliable, and repeatable results [3, 22, 49]. The last two years have witnessed an increase in the use of test automation across the software industry [24].

Test automation processes are still not mature in many companies [27]. Test maturity models have been used for a long time in the industry [27] and some models provide self-assessment instruments for organizations. For example, TOM [36] provides Test Organization Maturity Questionnaire, and TPI [34] provides Test Maturity Matrix. Among other things, self-assessment instruments enable identification of improvement areas and progress tracking if self-assessment is later repeated.

There are many reasons why valid self-assessment instruments of test automation maturity deserves more attention. First, organizations need to identify improvement steps at different stages of their test automation processes [27]. Current instruments for assessing the maturity of test automation processes are not sufficient. They lack coverage of test automation as a whole [29, 50, 51]. Second, we need to ensure that an instrument is valid to measure what it is supposed to measure [5, 21]. An invalid instrument may provide misleading information. Third, valid instruments may promote cooperative research efforts for empirical studies [14].

The objectives of this study are therefore to: (O1) synthesize what an organization should focus to assess its test automation processes; (O2) develop a self-assessment instrument (a survey) for assessing test automation processes and scientifically evaluate it.

To meet above objectives, we carried out the study in four stages. First, a literature review of 25 sources was conducted. Second, the initial instrument was developed. Third, seven experts from five Finnish and Swedish companies evaluated the initial instrument. Content Validity Index (CVI) [38, 39] and Cognitive Interview [23] methods were used. Fourth, the developed instrument was revised.

This study is part of TESTOMAT project (ITEA3), which proposes a Test Automation Improvement Model (TAIM) [19]. The instrument developed in this study is not an assessment method itself; rather, it is a component of TAIM, and that can be used to conduct the self-assessment for test automation processes.

The remainder of this paper is organized as follows: Section 2 presents the background and related work. Section 3 describes research methods. Section 4 presents results. Section 5 discusses findings and the implication. Section 6 presents threats to validity. Section 7 concludes this paper and states the future work.

## 2 BACKGROUND AND RELATED WORK

In this section, test maturity models and the aspects that they address about test automation are examined in Section 2.1. Self-assessment instruments are reviewed in Section 2.2, in order to observe the purposes for the use of those instruments.

### 2.1 Test maturity models

According to systematic literature review studies [4, 27, 50], TPI and its newer version TPI-Next [20], TMM [10], and TMMi [25] are the test maturity models widely used in the industry.

Test maturity models define key areas (KAs) that indicate where an organization should focus to assess its test process. The practices which are important to mature the test process are collected and mapped into different KAs [4, 11, 27, 46]. Table 1 presents example KAs of TestSPICE 3.0 with practices.

Table 2 outlines test maturity models that have KAs related to test automation. Upon further investigation, we found that the test maturity models address only some aspects of test automation but lack coverage of test automation as a whole.

TPI [34] defines 20 KAs. Test Tools is the only KA related to test automation. It assesses the extent to which test tools are used to support test activities, such as the planning and control, test specification, and test execution and analysis. TPI NEXT [20] inherits Test Tools KA from TPI.

TMM [10] indicates that the performance of test tools should be periodically evaluated, and automated test execution is done on the high maturity level. TMMi [25] was developed using TMM as the base. Several KAs of TMMi state the need to use test tools to support different test activities. For example, using test tools to support test design and execution practices.

TMap [33] has many KAs related to test automation. For example, Test Strategy KA describes process steps to develop a test automation strategy, Test Design KA describes solutions to create maintainable and reusable automated test cases, Test Policy KA defines guidelines for the use of test tools, etc.

PTMM [43] provides guidelines for cultivating practitioners to have the competency in software testing. It specifies what abilities and knowledge practitioners should have to perform different test automation activities.

**Table 1: Example KAs of TESTSPICE 3.0 ( [47] )**

| KAs | Practices |
|---|---|
| Test Automation Design | - Define potential test automation solutions for different test types and test stages<br>- Derive requirements for test automation tools<br>- Analyze constrains<br>- Decide on solutions<br>- Develop the detailed design for test automation solutions<br>- Perform proof of concept check of designed solution |
| Test Automation Usage | - Integrate test automation solution in the defined projects and test stages<br>- Run automated tests<br>- Interpret and classify test run results<br>- Report test run results |

**Table 2: Test maturity models**

| Model | Year | KA related to Test automation |
|---|---|---|
| TPI [34] | 1999 | Test Tools |
| TPI NEXT [20] | 2013 | Test Tools |
| TMM [10] | 1993 | Quality Control, Test Process Optimization |
| TMMi [25] | 2012 | No specific KA related to test automation. Test tools are considered as the complementary part of other KAs. |
| TMap [33] | 2014 | Test Strategy, Test Organization, Test Policy, Test Environments, Test Design, Test Tools, Test Professionals |
| PTMM [43] | 2006 | Test Roles, Test Skills |
| TestSPICE 3.0 [47] | 2014 | Test Planning, Test Environment Management, Test Data Management, Test Automation Process groups (including Test Automation Design, Test Automation Implementation, Test Automation Usage, etc.) |

TestSPICE 3.0 [47] sets indicators for assessing test automation maturity. For example, Test Environment Management and Test Data Management KA assess the extent to which the test environment is managaged and controlled for test automation.

### 2.2 Self-assessment instruments of test maturity models

Some test maturity models provide self-assessment instruments for organizations. For example, TPI provides a Test Maturity Matrix [34], TMap provides Checklists [1], TOM provides Test Organization Maturity Questionnaire [36]. The assessment items were defined in the form of survey questions. The assessment procedure is rather simple. Organizations answer assessment items (survey questions) to assess the maturity state of their test process [1, 34, 36].

The below example shows several assessment items taken from Test Environment Checklists of TMap [1]:

(1) Environmental data
- Are standard test data sets available?
- Do agreements about the test data exit with the test data owners?
- Does the system date need to be adapted?
- Is it possible to test with test accounts or with production profiles?
(2) Maintenance tools / processes
- Does one single point of contact exist for test environment maintenance?
- Are agreements reached about the readiness and quality of the test environment?
- Is the maintenance of the test environment supported by maintenance tools?

The purposes of the use of self-assessment instruments are to enable data collection, assist the assessment process, and conduct the self-assessment at different stages of the test process and therefore make it possible for the progress tracking and the identification of improvement steps [6, 36, 52].

## 3 RESEARCH METHOD

In software engineering, survey instruments are often developed in ad-hoc fashions [16, 32, 45], as Kitchenham and Pfleeger [32] described: "we often start from scratch, building models of a problem and designing survey instruments specifically for the problem at hand." Researchers (e.g.,[12, 17, 18]) have devoted the attention to instrument development issues in software engineering.

To design the research process of the study, we reviewed prior software engineering studies ([12, 18, 32]) that introduce main steps to develop an instrument and evaluate it for the validity. Furthermore, we learned from instrument development studies of other disciplines (e.g., [5, 31, 38]).

The research process of this study contains four stages named as literature review, instrument development, evaluation, and revision. The literature review stage addressed the objective (O1) of this study, see Section 1. The rest of stages addressed the objective (O2) of this study. Each stage is described in the following sections.

### 3.1 Literature review

We reviewed test maturity models to address the objective (O1) of this study. Our literature review was aided by a recent multi-vocal literature review [27], which includes both published and grey literature sources, on the test maturity assessment. This multi-vocal literature review identified 58 test maturity models that address test process assessment and improvement issues in the industry. We screened those models further against our criteria:

- Criterion #1: Is the model relevant to the scope of this study (the assessment of test automation maturity)?
- Criterion #2: Does the model contain practices of KA(s) related to test automation?
- Criterion #3: Was the model developed after 2004 (including 2004)?

Criterion #1 and #2 ensured that all content-relevant models are included to meet our literature review purposes. Based on our observation, the models developed before 2004 present very limited relevant results. Most of them did not have specific KAs related to test automation. Additionally, as this research domain is evolving rapidly, the earlier models may contain old technologies and processes. For example, many of them mainly focus on the test process that fits the traditional waterfall like software development. These methods have largely been replaced by agile methods since the introduction of the Agile Manifesto in 2001 [8]. The emergence and widespread use of new knowledge and technologies such as CI tools [13], Agile [15], DevOps [30], etc, also require the updates to test maturity models. Consequently, criterion #3 was defined.

Only 18 test maturity models [S1-S4, S6, S7, S9, S10, S14-20, S22-24] that meet all above criteria were finally selected for further reading.

We collected practices of KAs related to test automation in the chosen 18 models. Cruzes and Dybå's [12] thematic analysis principles were followed to qualitatively code data from those models. A predefined list of KAs (e.g., Test Automation Strategy, Test environment, Test design, Test execution, Measurements) was created according to KAs of TAIM, in order to classify the collected practices. A qualitative analysis software tool NVIVO [37] was used to code data from sources. During the process, we found that our predefined list of KAs was limiting, thus, the rest of practices of KAs collected from the sources, were coded by conducting 'inductive coding'. For example, additional codes were created according to the ISO 9001:2015 quality standard [2], aiming to include measurable quality attributes of test automation in the maturity assessment. At the end, by using the coded data, we collected practices mapped into 13 KAs of test automation.

Three academic (authors 2, 4, 7) and two industrial experts (authors 3, 5) reviewed the practices, which were mapped into 13 KAs of test automation. All academic experts have published in software test automation or extensively in software engineering. Two industrial experts have been working on software testing for decades and hold a relevant PhD degree. The collected practices in 13 KAs of test automation were shared with all reviewers through an online spreadsheet tool. The reviewers were asked to (1) review practices of KAs in relation to coded data from original sources, (2) give suggestions for the revision, (3) propose any new literature considered important to collect new practices those were not presented in our literature review. The commenting feature of an online spreadsheet tool was used to give comments and record details. Skype and face-to-face meetings were conducted in order to share opinions, discuss disagreements, and reach a consensus. As a result, we included seven additional literature sources [S5, S8, S11-S13, S21, S25] that contain practices of Test Environment, Test Design, Test Execution, and Verdicts KAs, which are important for test automation. We again coded those practices using NVivo. At the end, we collected practices mapped into 15 KAs of test automation.

### 3.2 Instrument development

In this stage, we developed the initial 77-item-instrument. Assessment items were created according to practices mapped into 15 KAs of test automation, as noted in preceding steps. We created one assessment item for each practice of a certain KA. The practices were translated into simple, easy-to-understand, and direct statements to form assessment items that could be related to everyday situations in the test process. Table 3 presents an example of the assessment item creation in Test execution KA. Respondents respond from 1 to 5 (from strongly disagree to strong agree) to indicate the degree of agreement with the statement of each assessment item.

**Table 3: Assessment item creation**

| |
|---|
| *A practice:* |
| Test cases are prioritized for the automation in order to meet the given schedule of test execution. |
| *is transformed into an assessment item:* |
| We prioritize test cases for the automation to meet the given schedule of test execution. |

## 3.3 Evaluation

We evaluate the content validity [5] of our initial 77-assessment-item instrument. Content Validity Index (CVI) [39] and Cognitive Interview [23] methods were used.

The content validity is the extent to which an instrument measures what it is intended to measure [31]. Researchers have developed rigorous methods in order to evaluate the content validity of the new instrument. One such method widely used is CVI [38, 39, 41, 44, 48]. CVI uses a team of experts to evaluate whether all assessment items of an instrument are relevant to its domain. The recommend number of experts is 5-10 [38]. The percentage of experts who agree on the relevance of each assessment item is calculated as I-CVI, and the average I-CVI across items is calculated as S-CVI/Ave. The content validity of each item is evaluated by I-CVI. The content validity of the entire instrument is evaluated by S-CVI/Ave. The studies [41, 42] suggested that, if three or more experts agree, an item with I-CVI$\geq$78 can be considered as having the excellent content validity, I-CVI $\leq$ .50 can be considered as having low content validity and deemed it is not acceptable, and .50 < I-CVI < .78 can be considered as having modest content validity. To ensure good content validity of the entire instrument, S-CVI/Ave should be .90 or higher [39].

To perform CVI evaluation, we selected seven test automation experts to review our instrument and evaluate the relevance of each assessment item on its domain - the assessment of test automation maturity. I-CVI and S-CVI/Ave were calculated to evaluate content validity of each item and the entire instrument. Cognitive interviews were conducted with selected experts to collect the data for CVI analysis. It is an interview technique that uses verbal 'probes' (open-ended questions) to specify information related to interview questions. The purposes of interviews in this study were to probe if experts understand each assessment item; observe how they rate each assessment item; explore new assessment items those are important but not appear in our instrument.

*3.3.1 Expert selection.* Seven test automation experts were selected from five Swedish and Finnish companies: Ericsson, Symbio, Comiq, Eficode, and a small software consulting company who wishes to remain anonymous. Those companies have various software related products or services, conducting a test automation process or offering test automation consulting services. Table 4 presents profile information of selected experts. Seven experts were labeled as Expert A-G in this study.

**Table 4: Profile information of experts**

| Education | four experts hold the Doctoral Degree<br>three experts hold the Master Degree |
|---|---|
| The current role in test automation | Test manager, Senior tester, QA engineer, Testing consultant |
| Years working on test automation | one expert with 8 years<br>five experts with 10-20 years<br>one expert with over 20 years |
| Company size | one small-size company (less than 50 employees)<br>three medium-size companies (50-249 employees)<br>one large-size company (more than 250 employees) |

*3.3.2 Data collection process.* We conducted interviews with the selected experts. Experts received our instrument in advance, so they could familiarize the content and prepare themselves for interviews. The interviews were conducted either face-to-face or via Skype. The duration was 85-150 minutes. The interview was audio recorded and notes were written down during the process.

During an interview, an expert answered each assessment item in our instrument. After that, he/she evaluated the relevance of each assessment item on the domain of our instrument - the assessment of test automation maturity. He/she evaluated by rating 1 = 'not at all relevant', 2 = 'not relevant', 3 = 'difficult to judge', 4 = 'relevant', 5 = 'highly relevant'. We prepared "probes" to dig specific information related to the rating. An expert was asked to explain: 'Can you understand this item', 'How did you evaluate', 'Do you have the suggestion to revise this item', 'Is this item difficult to rate for you'. Additionally, an expert was encouraged to point out new assessment items and give reasons.

*3.3.3 Data extraction and analysis.* The answers and ratings of all experts on each assessment item were collected with an online spreadsheet tool. Experts who rated 4 or 5 on a certain item were considered that they agree to its relevance on the domain of our instrument. We calculated I-CVI on each assessment item and S-CVI/Ave for the entire instrument. All assessment items were classified into three groups: excellent content validity items (I-CVI$\geq$.78), modest content validity items (.50 < I-CVI < .78) , and low content validity items (I-CVI$\leq$50). We played the audio recordings of interviews to explore how experts evaluate each assessment item. The narrative description of experts was transcribed verbatim.

We gathered new assessment items pointed out by experts in interviews. We compared the notes and audio recordings of interviews to observe why those are important to be included in the instrument of this study. The narrative description of experts was transcribed verbatim.

## 3.4 Revision

To revise our instrument for the better content validity, three authors (author 1, 2, 6) participated in the revision stage. We set a goal to achieve S-CVI/Ave.90 or higher, in order to ensure good content validity of the entire instrument. The results of evaluation stage were shared with all participants, including I-CVI distribution, the proposed new assessment items, transcribed data, etc. We discussed the changes with experts, who have already participated in our interviews. The commenting feature of an online spreadsheet tool was used to give the comments and record details. Skype and face-to-face meetings were conducted in order to share personal opinions, discuss disagreements, and reach a consensus.

## 4 RESULTS

The results of each stage are summarized in the following sections.

## 4.1 Literature review

The chosen sources in the pool are listed in Appendix A. Based on those sources, we collected practices mapped into 15 KAs that indicate where an organization should focus to assess its test automation process. Each KA is described below.

*4.1.1 Test Automation Strategy KA.* The test automation strategy defines strategic plans for test automation [S16, S20, S25]. The practices collected and mapped into this KA include:

- Test automation strategy that defines strategic plans for test automation is created. [S8, S11, S16, S20, S25]
- Unambiguous goals are set for test automation. [S5]
- A business case is created to conduct cost-benefit analysis of test automation.[S16, S24]
- The risks of test automation are identified and analyzed. [S10, S11, S16]
- Test scope will be automated to what degree is clearly defined. [S25]
- The overlap between automated testing and manual testing is considered. [S10, S11]
- The gaps and overlap between test types and levels are considered. [S16]
- Resources to perform test automation are identified, e.g., test tools, test data, test environment, skilled people. [S16, S24]
- Roles and responsibilities for test automation tasks are identified. [S16]
- The effort estimation for test automation tasks is made. [S10, S16]
- Feedback on the changes of test automation strategy are collected from stakeholders. [S16, S24]

*4.1.2 Resources KA.* Test automation processes are conducted with necessary resources [S6, S12, S16]. The practices collected and mapped into this KA include:

- There are enough skilled people (e.g.,experienced testers and managers, test consultants, automation experts) assembled to perform test automation. [S6, S11, S12, S25]
- There are enough funds to afford test automation. [S6, S11, S12, S25]
- There are enough time and effort allocated for test automation tasks. [S6, S11, S12, S25]
- There are enough test tools to support our testing activities. [S6, S11, S12, S25]
- Test environment in use is set up with all required software, hardware, test data. [S6, S11, S12, S25]

*4.1.3 Test organization KA.* Test organization is a organization unit, such as a test team, a department, or a whole organization, in where people are assembled to perform testing processes [S5, S8]. The practices collected and mapped into this KA include:

- keep test organization members motivated to perform test automation. [S10, S25]
- The roles and and responsibilities of test organization members are clearly defined. [S5]
- There are effective communication and problem solving mechanisms provided for test organization members. [S16, S20]
- There are the strong organizational support and management support for test automation. [S8, S17]
- Test organization has the enough expertise and technical skills to perform test automation, e.g., coding abilities [S8, S22], analysis skills [S8], domain knowledge (including system, product, etc) [S25], test design techniques [S10].
- Test organization has an insight of cost/profit ratio of test automation [S17]
- Test organization have the capability to choose suitable test tools for test automation tasks [S22]
- Test organization have the capability to to maintain test tools in use. [S16].

*4.1.4 Knowledge Transfer KA.* Test automation related knowledge is transferred within a company [S16]. The practices collected and mapped into this KA include:

- The expertise, good test automation practices, and good test tools are collected and shared for future projects. [S6, S16]
- Allow the time for training and the learning curve. [S17, S25]

*4.1.5 Test tool selection KA.* Selecting right test tools that best suit the needs is important for test automation [S5, S11, S17]. The practices collected and mapped into this KA include:

- The use of existing test tools is maximized. [S17]
- In test tool selection, the required features of test tools are identified , e.g., data-driven testing, scheduling, or mocking testing. [S4, S5, S9]
- In test tool selection, the important attributes of test tools are identified, e.g., usability, test code language, availability, costs, stability of test tools. [S4, S5, S8, S11, S14, S16, S18, S19]
- In test tool selection, the constrains to use test tools are identified, e.g., technical constraints of the test environment, funds, political issues. [S5]

*4.1.6 Test tool usage KA.* Test automaton relies on the use of test tools to support testing activities [S10, S16, S17]. The practices collected and mapped into this KA include:

- Preconditions to use test tools are elaborated, e.g., acquiring management commitment, understanding the modules in test tools, documents for maintenance. [S16]

- A business case to analyze Return on Investment of each test tool is created. [S10, S16, S17]
- There is a defined way to formally introduce new test tools within an organization unit, e.g., informing main users of test tools and collect feedback, raising the interest in new test tools. [S5, S17]
- New test tools are tested in pilot projects. [S8, S19]
- There is the board understanding about goals of using test tools. Test tools are evaluated periodically against those goals. [S3, S10, S16]
- There are the guidelines that define rules and principles for the use of test tools. [S7, S16]

*4.1.7 Test Environment KA.* Test environment set up with software, hardware, test data, etc., is managed and controlled to execute automated tests [S25, S21]. The practices collected and mapped into this KA include:

- The requirements on test environment are thoroughly understood. [S24]
- The configuration of test environment is under the control. [S21]
- Test environment and test data are tested before it is ready to use. [S25]
- Test environment support is provided to aid the test automation construction. [S25]
- Test environment fault or unique aspects like timing delays, or configuration variants are promptly identified. [S25]

*4.1.8 Test Requirements KA.* Test automation requirements need to be identified and derived [S24]. The practices collected and mapped into this KA include:

- There is the defined way to derive test automation requirements. [S8, S24]
- The changes that affect test automation requirements are controlled. [S24]

*4.1.9 Test Design KA.* The test design is about use techniques to create test cases for test automation [S04, S17, S24]. The practices collected and mapped into this KA include:

- There is the specific test design techniques to create test cases. [S4, S17]
- The patterns in the use of test design techniques are captured and reused. [S04]
- Test cases are selected into test suites for different purposes, e.g., smoke, regression, field testing. [S8]
- There are the guidelines on designing test cases, e.g., coding standards, test-data handling methods, specific test design techniques, processes for reporting and storing test results. [S4, S5, S16, S24]
- Static and dynamic measurements are performed on test code. [S4]

*4.1.10 Test Execution KA.* Test execution refers to processes that automated test cases/suites are actually executed [S5, S20]. The practices collected and mapped into this KA include:

- Test cases are prioritized for the automation in order to meet the given schedule of test execution. [S5, S20]

- Pre-processing tasks are automatically performed before the execution, e.g., automatically create required files, database or data, recognize files or data, and convert test data into the required format [S5].
- Post-processing tasks are automatically performed after the execution, e.g., automatically delete files or database records, and convert outcomes into the required form [S5].

*4.1.11 Verdicts KA.* Verdicts indicate the result of executing an automated test case [S9, S16, S24]. They are generated and collected from test execution. The practices collected and mapped into this KA include:

- There are stable and predictable test oracles to determine how a system behaves when it pass or fail a test [S8].
- The status and progress of testing can be monitored with respect to test results [S8].
- Test results collected from different sources are managed and integrated into a big picture.[S9, S16, S24]
- Useful test results are reported to relevant stakeholders [S16].
- The dashboard is adapted to each stakeholder [S25].

*4.1.12 Test Automation Process KA.* Test automation process defines the approach to conduct test automation in the test process [S3, S7, S10, S16]. The practices collected and mapped into this KA include:

- Test automation is conducted in the stable and controllable test process. [S3, S7, S10, S16, S25]
- Test automation is conducted in parallel to development cycles. [S3]
- Test automation process is built to support other processes, e.g., software development, software deployment, software maintenance, and the business as the whole. [S2, S16]

*4.1.13 Software Under Test KA.* Software Under Test (SUT) related factors have the impact on test automation maturity [S5, S8]. The practices collected and mapped into this KA include:

- SUT is mature enough to conduct test automation [S8].
- SUT is testable by test automation. [S5, S16, S17]
- Execution speed of SUT is high enough to conduct test automation [S5].

*4.1.14 Measurements KA.* Measurements are quantified observations for the quality and/or performance of test automation [S5, S16]. They are used to measure, control, and track the test process for improvement steps. The practices collected and mapped into this KA include:

- The right metrics were used to measure test automation.[S5, S16, S17]
- Important attributes of test automation are identified, e.g., test effectiveness, test thoroughness, efficiency, reliability, maintainability [S5].
- Improvements areas are identified through measurements. [S5, S16, S17]
- Test organization members frequently get feedback about their performance. [S5, S16]

*4.1.15 Quality Attributes KA.* This KA presents measurable attributes of test automation to conduct measures for its quality [S5]. We found 6 attributes of test automation can be measured, as described below:

(1) Portability:
- The ease of running automated tests in a new environment with different hardware, software environment, configurations, etc [S5].

(2) Maintainability:
- The extent to which testware (e.g., test cases, test data, test results, test reports, expected outcomes and other artifacts generated for automated tests) is organized in the good architecture. [S5, S10, S15, S16]
- The ease of managing (e.g., 'keep it alive', provide services or user support, fix bugs) and updating (e.g., the deployment and development) test environment. [S21, S24, S25]
- The ease of maintaining automated tests and keep them operational [S5].

(3) Reliability:
- The extent to which reliable and accurate results are produced by automated tests [S5].
- The extent to which automated tests are resistant to inconsequential changes (Software Under Test changes, requirement changes, unexpected events in test execution, etc). [S5, S8]
- Test environment has the high accessibility for executing automated tests. [S21, S25]
- Restoration and recovery mechanisms are built to enable the test environment back to the previous status. [S21]

(4) Usability:
- The ease of using automated tests by different types of users such as testers, managers, and leaderships [S5].
- The ease of use of test environment [S5].

(5) Efficiency:
- The extent to which Automated tests are conducted with the estimated costs and effort. [S5, S16]

(6) Functionality:
- Automated tests meet the given test purposes and bring the benefits, e.g., better detection of defects, increasing test coverage, reducing test cycles, good Return on Investment, better product quality. [S1, S5, S11, S13]

## 4.2 The developed instrument

We created the initial 77-assessment-item instrument, which is presented in: https://figshare.com/s/20aeb06772f0136e627b.

## 4.3 Evaluation

According to answers of experts on assessment items, we found that our assessment items are prone to social desirability response bias where respondents deny undesirable answers and give responses that are more (socially) desirable in their work context. For example, experts who are managers/leaders in test automation, tend to give more positive answers and discuss less about the impediments that they face in their test automation process.

The distribution of assessment items in three different content validity groups is presented in Table 5. Of 77 initial assessment items, 68 had the excellent content validity evidence (I-CVI ≥ .78), 4 had the modest content validity evidence (.50 < I-CVI < .78), 5 had the low content validity evidence (I-CVI ≤ 50). S-CVI/Ave.84 indicates that the initial instrument have the high percentage of valid assessment items for assessing test automation maturity, but there is a space to improve content validity of the entire instrument.

**Table 5: The distribution of assessment items**

| Dimension | f: Excellent | f: Modest | f: Low | Total |
|---|---|---|---|---|
| Test Automation Strategy | 10 | 0 | 1 | 11 |
| Resources | 5 | 0 | 0 | 5 |
| Test Organization | 6 | 0 | 2 | 8 |
| Knowledge Transfer | 2 | 0 | 0 | 2 |
| Test Tool Selection | 3 | 0 | 1 | 4 |
| Test Tool Usage | 5 | 1 | 0 | 6 |
| Test Environment | 5 | 0 | 0 | 5 |
| Test Requirements | 2 | 0 | 0 | 2 |
| Test Design | 4 | 1 | 0 | 5 |
| Test Execution | 3 | 0 | 0 | 3 |
| Verdicts | 5 | 0 | 0 | 5 |
| Test Automation Process | 3 | 0 | 0 | 3 |
| Software Under Test | 3 | 0 | 0 | 3 |
| Measurements | 4 | 0 | 0 | 4 |
| Quality Attributes | 8 | 2 | 1 | 11 |
| Total | 68 | 4 | 5 | 77 |
| S-CVI/Ave .84 | | | | |

All assessment items in Resources, Knowledge Transfer, Test Environment, Test Requirements, Test Automation Process, Software Under Test, Test Execution, Verdicts, Measurements KAs had the excellent content validity evidence (I-CVI ≥.78).

Table 6 lists modest content validity Items. Experts proposed suggestions to revise those items:

- SQ36: need examples to explain 'rules and principles for test tool usage.' *(Expert A, F, G)*
- SQ44: need to explain what 'test design techniques' should be specified. *(Expert A, C, F, G)*
- SQ67: not all automated tests need the high portability. *(Expert A, E)*
- SQ75: need examples to explain 'different types of users for test environment'. *(Expert B)*

Table 7 lists low content validity items. The reasons of why those items are invalid are summarized below:

- SQ7: No empirical evidence to prove this will contribute to test automation maturity *(Expert B)*; This is more suitable for traditional software developments, but difficult to conduct for agile developments *(Expert D, G)*.
- SQ22, SO23: The current industry not emphasize that a test organization should have such capabilities, skills, or abilities. *(Expert A, B, C, D, E, F, G)*
- SQ27: no empirical evidence to prove this will contribute to test automation maturity *(Expert B, C)*. This item is repeated with strategic planning items that ask if a company identify resources including test tools *(Expert A, D, G)*.

### Table 6: Modest content validity items

| Assessment items | I-CVI |
|---|---|
| SQ36. We have the guidelines that define the rules and principles for test tool usage. | .71 |
| SQ44. We use specific test design techniques to create test cases. | .71 |
| SQ67. Our automated tests are easy to be set up in a new environment with different hardware, software environments, configurations, etc. | .57 |
| SQ75. Our test environment has the high usability for different types of users. | .71 |

### Table 7: Low content validity items

| Assessment items | I-CVI |
|---|---|
| SQ7. We consider the gaps and overlaps between test types and levels. | .43 |
| SQ22. Our test organization has an insight into the cost/profit ratio of our test tools. | .14 |
| SQ23. Our test organization is capable to choose the most suitable test tools for their tasks. | .29 |
| SQ27. We check if the exiting test tools can meet our needs before buying the new ones. | .29 |
| SQ76. Our automated tests have the high usability for different types of users such as testers, managers, and leaderships. | .43 |

- SQ76: It is not a measurable quality attribute in the test process. *(Expert A, D, E, G)*

Experts pointed out 8 new assessment items, which are important to be added into our instrument:

- NSQ1. We use test data in compliance with the industry regulations, legislation and the baseline of the undertaken project. *(Expert G)*
- NSQ2. We are capable to manage (e.g., generate, analyze, and synthesize) test data correctly. *(Expert B & G)*
- NSQ3. Our test environment matches production environment well. *(Expert F)*
- NSQ4. We conduct parallel execution for the complex system. *(Expert D & G)*
- NSQ5. We have notifications to alarm the critical failures of tests. *(Expert F)*
- NSQ6. Our measurements are visible, e.g., as part of test reports or shown on dashboards. *(Expert D & G )*
- NSQ7. We have fast feedback cycles for test automation development. *(Expert E & F)*
- NSQ8. Our test automation strategy is a living document that is periodically reviewed and updated depending on our present needs *(Expert E & F)*

## 4.4 The Revision

We made the revision to our instrument, as shown in Table 8. In total, 4 modest content validity items were modified, 5 low content validity items were deleted, 8 new items were added. As the result, S-CVI/Ave was increased to .91, which meets the standard (S-CVI/Ave .90) of good content validity for the entire instrument. At the end, a final 80-assessment-item instrument was developed and it was presented in: https://figshare.com/s/ad189d406e48b32e23d4.

### Table 8: The revision

| KA | Revision |
|---|---|
| Test Automation Strategy | delete SQ7 |
|  | add NSQ8 |
| Test Organization | delete SQ22, SQ23 |
| Test Tool Selection | delete SQ27 |
| Test Tool Usage | modify SQ36 |
| Test Environment | add NSQ1, NSQ2, NSQ3 |
| Test Design | modify SQ44 |
| Test execution | add NSQ4, NSQ5 |
| Test automation process | add NSQ7 |
| Measurements | add NSQ6 |
| Quality attributes | modify SQ67, SQ75 |
|  | delete SQ76 |

## 5 DISCUSSION

We had many thought-provoking discussion about the findings and implications of some research stages, as discussed below.

### 5.1 Instrument development in software engineering

As described in Section 3, survey instrument development issues have been inadequately addressed in software engineering. In many other disciplines, such as social science [7], nursing research [38, 41], and education [39], researchers usually pay greater attention on instrument development issues. They follow rigorous guidelines to develop a new instrument and evaluate it for the validity [38, 41].

In the study of this paper, we learn from prior software engineering studies, as well as relevant studies of other disciplines to develop and evaluate the instrument. The development and evaluation process was documented in this paper, in order to raise the attention to instrument development in software engineering.

### 5.2 Instrument Development Considerations

We debated several topics in the instrument development stage, and those are summarized there.

First, the "cost" of answering assessment items was debated. Based on the interviews, it took 20-40 minutes to answer all assessment items. Normally, for respondents, "satisfying" [35] is a strategy that minimize the effort to answer large numbers of assessment items. In this case, they tend to provide answers to get rid of work rather than right answers. The acceptable response time is likely to be context dependent. One extreme practitioner might say that the acceptable response time to answer a survey is 10 seconds. On the other extreme CMM assessments [28], in the 1990's, could take up to 5 days on site and it has a survey [52] with 117 questions as only one part of the assessment. Consequently, it is necessary to use strategies to handle satisfying behavior, in order to develop instruments with the different acceptable response time, e.g., we could have for instance 10, 20, 40 and full 80 -assessment-item versions. Similarly, we need to produce versions with different assessment items for different roles. For example, we can have separate assessment items for management, the users of test automation, experts who developing the test automation.

Second, we debated response bias and ways to deal with it. Response bias has been a research topic in psychology and sociology for decades [26, 40]. It presents in self-reported data and means the tendency of a respondent answer incorrectly [26]. In particular, as described in 4.3, our assessment items are prone to social desirability response bias. In the next step, we will investigate techniques to control response bias from psychology [40]. Additionally, we will work towards automating or getting objective measures on test maturity to help in controlling response bias.

### 5.3 Instrument evaluation

In the evaluation stage of this study, data saturation was achieved after conducting five interviews. The responses of experts about how they rate the content validity of each assessment item started to be repetitive, and there were no new assessment items pointed out afterwards. This indicates the appropriateness and adequacy was achieved with the current sample size.

## 6 THREATS TO VALIDITY

Criterion validity [5, 31] refers to the extent to which the instrument is related to the other measures of the relevant criterion. As our instrument in this study is purported to assess test automation maturity for improvement steps, predictive validity can be investigated. We plan to pilot this instrument in several companies at the different stages of our project, in order to examine the assessment results and later effects on test automation process in those companies. However, the study of this paper is the first attempt to develop a self-assessment instrument for test automation maturity. Comparing assessment results of our instrument with the ones of other independent instrument suffers the difficulty.

Construct validity [5, 31] concerns about how well an instrument can measure what it is supposed to measure. All other types of validity evidence, including content validity and criterion validity, can make contributions to construct validity. However, it may need the following studies to empirically evaluating construct validity of our instrument in this study.

## 7 CONCLUSION

This study made three main contributions. First, it collected practices mapped into 15 KAs that indicate where an organization should focus to assess its test automation maturity. This will help companies and researchers to better understand test automation practices and processes. Second, our self-assessment instrument was developed and evaluated using scientific methods. The development and evaluation process is demonstrated. As noted in Section 3 and 5.1, survey instruments are often developed in ad-hoc fashions in software engineering. We argue that our work could act as an example on how to create assessment instruments also for other areas of software engineering than just test automation. Third, we discuss important topics such as response bias that threatens self-assessment instruments and the cost of answering the survey.

In the future, we plan to map practices of KAs into the maturity levels of TAIM model, and establish a benchmark for companies to compare themselves with the rest of industry. A number of people will be invited to use our instrument to conduct self-assessment for their test automation processes. The assessment data will be entered into our database for tracking the progress of each KA in their test automation. A the same time, the criterion validity, construct validity, and reliability of our instrument will be evaluated. Additionally, as noted in Section 5, it is necessary to address the acceptable responding time, and find ways to counter response bias, when distributing our instrument.

## ACKNOWLEDGMENTS

The authors would like to thank companies and individuals who participated in interviews. This work has been supported by TESTOMAT Project (ITEA3 ID number 16032), funded by Business Finland under Grant Decision ID 3192/31/2017.

## APPENDIX A. A POOL OF SOURCES